# Teaching Using Immersion - Explaining Magnetism and Eclipses in a Planetarium Dome

Patricia H Reiff [1] and Carolyn Sumners [2]

1. Rice University Department of Physics and Astronomy, 6100 Main St MS 108, Houston TX 77005; reiff@rice.edu
2. Houston Museum of Natural Science, 5555 Herman Park Dr., Houston TX 77004; csumners@hmns.org

## Abstract:

Previously we have shown that three-dimensional concepts are more readily learned in a three-dimensional context. Although VR headsets are growing in popularity, they only provide a quite limited field of view, and each person in a group may be viewing a different direction or a different time in the visualization. By using instead a fullsphere movie (VR360) in a planetarium dome instead of a headset, you can "share the VR"® and specify which half of the sphere your audience is looking at. You can pause the movie, ask questions using a clicker system, display the results, and move on if the subject is mastered or explain more if items are not understood. This paper shows the results of teaching magnetism in a dome theater, showing that both students and teachers nearly double their understanding of magnetism topics after one viewing. We also created seven animations explaining eclipses that were distributed free to nearly 200 planetariums. Listing of concepts learned by teachers in our live eclipse program are shown.

**Background**

In this paper we have used a planetarium dome in its more traditional "hemisphere" mode to teach about magnetism (using our new show "Magnetism – Defending Our Planet, Defining the Cosmos") and pre/post testing to show how many concepts can be understood in a relatively short experience. We had previously shown [Sumners et al., 2008] that dome learning is more effective on three dimensional concepts and that dome learning leads to longer content retention [Zimmerman et al., 2014]. We have identified 35 concepts that most high school students do NOT know about magnetism, and have done pre/post testing on students and teachers. Most students more than doubled the number of concepts that they were able to explain after watching our "Magnetism" show [Sumners and Reiff, 2017] just one time. The flatscreen version of the show can be watched in its entirety at this website: http://www.eplanetarium.com/shows/ddome/rice/magnetism/. These are preliminary results which will be expanded in the next year:

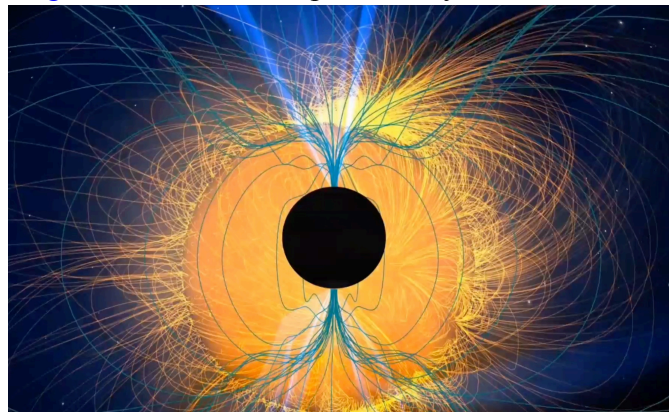

**Magnetism Topics tested:**
1. What force accelerates particles in a super collider?
2. What can we learn from the paths of charged particles after a collision?
3. What force makes particles rise up from the sun's surface in arcs?

*Figure 1. Exoplanet magnetic field (from "Magnetism" show)*

4. What force makes these particles return to the sun?
5. Where do charged particles enter Earth's atmosphere?
6. What causes auroras?
7. What causes their colors?
8. Describe different auroral shapes.
9. What element causes the Earth to have a magnetic field?
10. What properties are required for a planet to have a magnetic field?
11. How do sharks and some other animals use magnetism to navigate?
12. How has the Earth's magnetic field changed over time?
13. How do we know it has changed?
14. Why should we worry about the Earth's changing magnetic field?
15. Who might be affected?
16. What is the MMS mission?
17. What's special about the MMS spacecraft?
18. What happens at reconnection sites?
19. Why is it important for a planet to have a magnetic field?
20. When are astronauts beyond Earth's magnetic field?
21. Which inner planets have magnetic fields? Which do not?
22. Why don't these planets have magnetic fields?
23. Under what conditions did Mars have surface water?
    Does Mars have surface water now?
24. Does Jupiter have a magnetic field? If it does, how strong is it?
25. What compounds are in Uranus's atmosphere?
27. How is the aurora on a brown dwarf different from an Earth aurora?
28. About how many extra-solar planets have we discovered?
29. How did we detect the magnetic field around another planet?
30. About how many stars are in our Milky Way Galaxy?
31. Does the Milky Way have a magnetic field?
32. How can we detect it?
33. What is the most energetic electromagnetic radiation and how do we detect it?
34. How do magnetic fields affect the core of a supernova remnant?
35. What would we know if we found auroras around a distant planet?

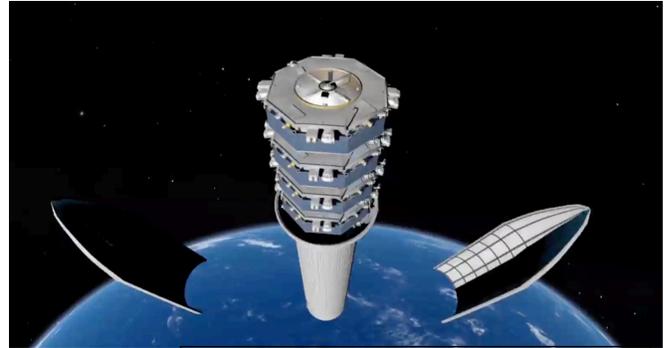

*Figure 2. MMS spacecraft (from "Magnetism" show)*

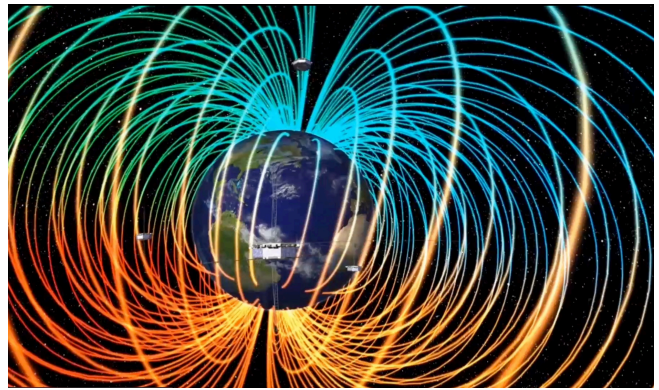

*Figure 3. MMS in orbit (from "Magnetism" show)*

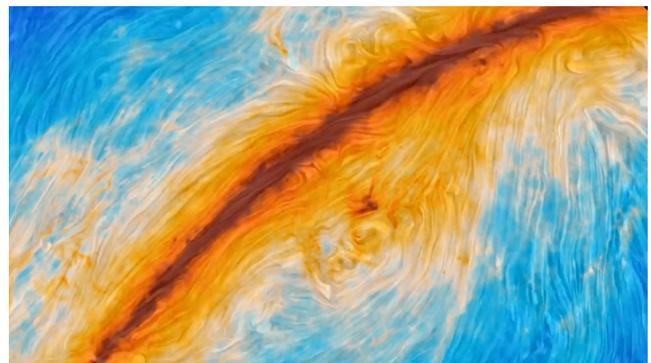

*Figure 4. Galactic magnetic field (from "Magnetism" show)*

**Table 1. Results**

| Tested Group | Number answered pretest | Number answered after one viewing | Number answered after two viewings | Gain |
|---|---|---|---|---|
| Science students entering High School Senior year with one year of Chemistry and Physics | 0 | 21 | 30 | 21 |
| Science students entering their HS Senior Year who worked on the Magnetism show in the summer of 2016 and had had Chemistry and Physics | 17 | 34 | | 17 |
| One student entering HS junior year who had Chemistry but not Physics | 13 | 26 | | 13 |
| One student entering HS freshman who had not taken Chemistry or Physics | 5 | 20 | | 15 |
| Science Teachers (5) took multiple choice version of the test<br>Standard deviation, confidence level | 17.6 +/- 4.7 | 25 +/- 4 | | 7.4 (.1%) |

**Eclipse Animations Released**

We have also created a series of seven eclipse animations to teach about solar and lunar eclipses. These animations have been used in more than 200 planetarium theaters and used as part of several TV specials on the August 2017 eclipse. By teaching eclipses in a dome, the students correctly understand the three-dimensional geometry of the Earth and Moon orbits and the causes of eclipses.  Preview and low-resolution fisheye versions are available free here: http://space.rice.edu/eclipse/eclipse_animations.html

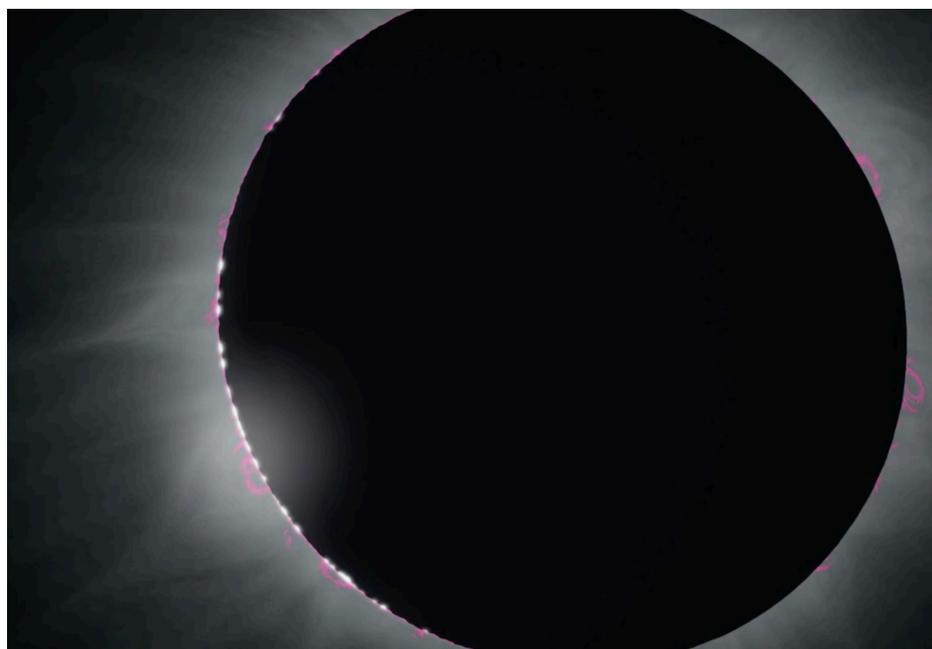

*Figure 5.  Baily's beads, from the most popular of the eclipse animations (based on number of times shown on national TV and Youtube viewings)*

**Estimated number of people reached:**
  In person or web training, countable:  5,101  (Includes public outreach events)
  Via Youtube:  43,348
  Via Planetariums:  ~0.1 - 1 million?
  Via News (TV and radio broadcast, web interviews):  ???
  Via Dome Loaner program: estimated 4,000
  On our email server:  (some people signed up for more that one segment)
        Planetarium educators (166)
        Informal educators (122)
        K-12 Educators (498)
        Higher Education Educators (101)
        Media (31)

A fisheye time-lapse of the eclipse from near Dubois, Wyoming is here, along with other images and videos from space:
http://space.rice.edu/eclipse/eclipse_2017.html

We gave in person, online live, and online prerecorded versions of our Eclipse training.  This training included the videos we produced, plus additional material in a powerpoint.  To receive Professional Development hours, the teacher who watched the prerecorded version had to send in an email listing four things they learned.  They could receive more hours by doing homework on apparent sizes of the Sun and Moon, and ellipticity.
Here are the most popular topics learned, among the 44 responses we received.  Note that of the top 9 topics, six were learned from our seven eclipse animations, and the others from the powerpoint.

| *Topics mentioned as learned from training (44 teachers listed four or more topics they learned)* | Number of Teachers who said they learned this topic | Where learned |
|---|---|---|
| Baily's beads and Diamond Ring | 15 | Video |
| How to safely watch an eclipse | 15 | Powerpoint |
| Penumbra is partial; umbra is total | 13 | PP Diagram |
| USA only country experiencing totality | 12 | Video, PP |
| Selenelion is lunar eclipse at dawn or dusk | 11 | Powerpoint |
| Eclipse has four contacts | 10 | Video |
| Corona only visible during totality | 10 | Video |
| Difference between lunar and solar eclipses | 10 | Video |
| Only safe to use your naked eyes during totality | 10 | Video |
| Binocular tips (solar filters; tripod adapter; each person should have one for totality) | 9 | Powerpoint |
| Shadow moving 1000 mph from NW | 8 | Powerpoint |
| Punch a pinhole image to shoot the shadow | 8 | Powerpoint |
| Annular eclipse umbra doesn't reach Earth | 6 | PP Diagram |
| 1.5 hours for the shadow to cross US | 5 | Video, PP |
| Take an hour to cover the Sun | 5 | Video, PP |
| Clouds become more transparent | 5 | Powerpoint |
| Use Google map to get timings, coverage | 4 | Powerpoint |
| Parts of the Sun (prominences, streamers) | 3 | Powerpoint |
| ***Topics learned from Homework (not required)*** | | |
| Angular size of Moon versus Sun | 3 | Homework |
| Moon's apparent size varies | 3 | Homework |
| Earth is 4x diameter of Moon | 1 | Homework |
| Moon is receding from Earth | 1 | Homework |

## Summary and Conclusions

Planetarium animations are an effective tool to teach about magnetism and the solar eclipse. The magnetism study is ongoing and will be improved with a larger set of subjects and questions.

## References

Sumners, C. and P. H. Reiff, "Magnetism", planetarium show, Evans and Sutherland, 2017.

Sumners, C., P. H. Reiff, and W. Weber, "Learning in an Interactive Digital Theater," *Advances in Space Research*, DOI:10.1016/j.asr.2008.06.018, Vol 42, p. 1848-1854 (2008).

Zimmerman, L., S. Spillane, P. Reiff, and C. Sumners, Comparison of Student Learning about Space in Immersive and Computer Environments, *Journal and Review of Astronomy Education and Outreach*, *V1*, p. A5-A20, (2014).